\documentclass[12pt,preprint]{aastex}

\usepackage{epsfig}
                                                                                
\def\vkm{km s$^{-1}$}
\def\vkme{\textrm{km s}^{-1}}
\def\degree{$^\circ$}

\def\arcsa#1#2{$#1^{\prime\prime}_{^\textrm{.}}#2$}

\def\smassrate{$M_\odot$ yr$^{-1}$}

\def\Jupmass{$M_\textrm{\scriptsize Jup}$}

\def\Jyb{Jy beam$^{-1}$}

\def\Jybk{Jy beam$^{-1}$ km s$^{-1}$}

\def\tlabel#1{(\textit{#1})}

\def\cmc{cm$^{-3}$}
\def\cms{cm$^{-2}$}

\def\eqt#1#2{#1_\textrm{\scriptsize #2}}

\def\nH{n_\textrm{\scriptsize H}}

\def\SOt{$N_J=8_9-7_8$}
\def\H2{H$_2$}
\def\N2HP{N$_2$H$^+$}

\def\NH3{NH$_3$}

\def\COt{$J=3-2$}
\def\SiOt{$J=8-7$}
\def\SOt{$N_J=8_9-7_8$}

\def\HCOP{HCO$^+$}

\def\putfig#1#2#3{\epsfig{scale=#1,angle=#2,figure=#3}}
\def\leftblank#1{}

\begin{document}

\title{The reflection-symmetric wiggle of the young protostellar jet HH 211}
\author{Chin-Fei Lee\altaffilmark{1},
Tatsuhiko I. Hasegawa\altaffilmark{1},
Naomi Hirano\altaffilmark{1},
Aina Palau\altaffilmark{2},
Hsien Shang\altaffilmark{1}, 
Paul T.P. Ho\altaffilmark{1,3}, and
Qizhou Zhang\altaffilmark{3}
}
\altaffiltext{1}{Academia Sinica Institute of Astronomy and Astrophysics,
P.O. Box 23-141, Taipei 106, Taiwan; cflee@asiaa.sinica.edu.tw}
\altaffiltext{2}{
Laboratorio de Astrof\'{\i}sica Estelar y Exoplanetas, Centro de                                          
Astrobiolog\'{\i}a (INTA-CSIC),                                                                         
LAEFF campus, P.O. Box 78, E-28691 Villanueva de la Ca\~nada (Madrid), Spain}
\altaffiltext{3}{Harvard-Smithsonian Center for Astrophysics, 60 Garden
Street, Cambridge, MA 02138}

\begin{abstract} 
HH 211 is a highly collimated jet originating from a nearby young Class 0
protostar. Here is a follow-up study of the jet with our previous
observations at unprecedented resolution up to $\sim$ \arcsa{0}{3} in SiO
(\SiOt{}), CO (\COt{}), and SO (\SOt{}). SiO, CO, and SO can all be a good
tracer of the HH 211 jet, tracing the internal shocks in the jet.
Although the emissions of these molecules show roughly the same morphology
of the jet, there are detailed differences. In particular, the CO emission
traces the jet closer to the source than the SiO and SO emissions. In
addition, in the better resolved internal shocks, both the CO and SO
emission are seen slightly ahead of the SiO emission. The jet is clearly
seen on both sides of the source with more than one cycle of wiggle. The
wiggle is reflection-symmetric about the source and can be reasonably fitted
by an orbiting source jet model. The best-fit parameters suggest that the
source itself could be a very low-mass protobinary with a total mass of
$\sim$ 60 \Jupmass{} and a binary separation of $\sim$ 4.6 AU. The
abundances of SiO and SO in the gas phase are found to be highly enhanced in
the jet as compared to the quiescent molecular clouds, even close to within
300 AU from the source where the dynamical time scale is $<10$ yrs. The
abundance enhancements of these molecules are closely related to the
internal shocks. 
The detected SiO is either the consequence of
the release of Si-bearing material
from dust grains or of its formation via gas chemistry in the shocks.
The SO, on the other hand,
seems to form via gas chemistry in the shocks.
\end{abstract}


\section{Introduction}

Jets from protostars represent one of the most intriguing signposts of star
formation. They are believed to be launched from accretion disks around the
protostars. They are highly supersonic, collimated, and ballistic. Yet, they
are often found to show wiggles in their trajectories. These wiggles of the
jets could be due to a precession of the jets \cite[e.g.,][]{Raga1993}, or
an orbital motion of the jet sources (i.e, the protostars that drive the
jets) \cite[e.g.,][]{Fendt1998,Masciadri2002}, or both. The jet precession
may be due to a precession of the accretion disks 
because of tidal interactions in noncoplanar binary systems
\cite[see, e.g,][]{Terquem1999} and it
can give rise to point-symmetric (i.e., S-shaped) wiggles, while the orbital
motion of the jet sources may arise in binaries and it can give rise to
reflection-symmetric (i.e., C-shaped) wiggles \citep{Fendt1998}.
\citet{Masciadri2002} have studied the wiggles of the DG Tau microjet, HH
47, and the Serpens radio continuum jet. Similarly, \citet{Anglada2007} have
studied the wiggle of the HH 30 jet. In both cases, assuming that the wiggles of those jets
are due to an orbital motion of the jet sources, they have derived orbital
parameters and masses that are reasonable for pre-main-sequence binaries.
For those jets, however, only one side of the jets is seen, and thus we can
not confirm if the wiggles are really reflection-symmetric.


This paper is a follow-up study of the HH 211 jet with our previous
observations at unprecedented resolution up to $\sim$ \arcsa{0}{3} (Lee et
al. 2009, hereafter \citet{Lee2009HH211}), obtained with the Submillimeter
Array\footnote{The Submillimeter Array is a joint project between the
Smithsonian Astrophysical Observatory and the Academia Sinica Institute of
Astronomy and Astrophysics, and is funded by the Smithsonian Institution and
the Academia Sinica.} \citep{Ho2004}. The jet is clearly seen in SiO
(\SiOt{}) on both sides of a young Class 0 source with more than one cycle
of wiggle, and is thus one of the best candidates to study the origin of the
wiggle.  The wiggle is reflection-symmetric and it seems to be due to an
orbital motion of the jet source \citep{Lee2009HH211}. The wiggle was called
the C-shaped bending in \citet{Lee2009HH211} but not here anymore because
more than one cycle is seen. In order to have a more complete picture of the
wiggle, here we also present CO (\COt{}) and SO (\SOt{}) emissions of the
jet in the same observations as already described in \citet{Lee2009HH211}.
By modeling the wiggle, we derive the orbital parameters and then discuss
the nature of the source. In addition, we also
refine the mass-loss rate of the jet with the CO emission at high resolution
and investigate the possible origins of the SiO, SO, and CO emissions in the
jet.

\section{Results}

In the following, we present the morphology and kinematics of the jet in CO
(\COt{}) and SO (\SOt{}) emissions in comparison to those in SiO (\SiOt{})
emission in \citet{Lee2009HH211}. We also derive the density of the jet from
the CO emission and then the abundances of SO and SiO. 
The systemic velocity in HH 211 is assumed to be $9.2\pm0.3$ \vkm{} LSR and 
the distance to the HH 211 is assumed to be 280$\pm40$ pc, as in
\citet{Lee2009HH211}. The jet is almost in the plane of the sky with an
inclination angle of $\sim$ 5\degree{}$\pm3$\degree{}. The eastern component of the jet is blueshifted while
the western component is redshifted. The eastern component of the jet has a
P.A. of 116.1\degree{}$\pm0.5$\degree{}, while the western component has a
P.A. of 297.1\degree{}$\pm0.5$\degree{}, indicating that the jet has a
small-scale bending, $\sim$ 0.5\degree{} to the northeast
\citep{Lee2009HH211}.

\subsection{Morphology of the jet} 

Figure \ref{fig:jet} shows the CO and SO maps on top of the SiO
map of the jet and the
352 GHz continuum map (green contours) of the envelope-disk from
\citet{Lee2009HH211}.
The CO emission of the jet is derived by excluding the CO emission within
$\sim$ $\pm$ 10 \vkm{} from the systemic, in order to avoid the contamination from
the shells and internal shells \cite[for more details, see][]{Lee2007HH211}.
The CO emission of the jet is more than a factor of 4 weaker than the SiO
emission. Although the CO emission shows roughly the same morphology as the
SiO emission, there are detailed differences (Fig. \ref{fig:jet}b). The CO
emission shows a chain of paired knots on either side of the source SMM1,
but with the peaks slightly ahead (i.e., downstream) of the SiO peaks for
most of the better resolved knots (see, e.g., knots BK2, BK3, and RK2 in
Fig. \ref{fig:jet}c). The CO emission also shows the same curvy structures
for knots BK1 and RK1, and the same reflection-symmetric wiggle of the jet
(Figs. \ref{fig:jet}b, c) as SiO. However, the CO emission extends closer in
to the source than the SiO emission, with the closest peaks (labeled as
knots RK0 and BK0 in Fig. \ref{fig:jet}c) at $\sim$ \arcsa{0}{4} ($\sim$ 100
AU) on either side of the source. The SO emission is even weaker than the CO
emission, but it also shows knotty structures along the jet, like the CO
emission (Fig. \ref{fig:jet}a). Like the CO peaks, the SO peaks also appear
slightly ahead of the SiO peaks in e.g., knots BK2, BK3, and RK2. However,
almost no SO emission is seen toward the innermost pair of CO knots.

\subsection{Kinematics along the jet axis}\label{sec:kinematics}

\begin{figure} [!hbp]
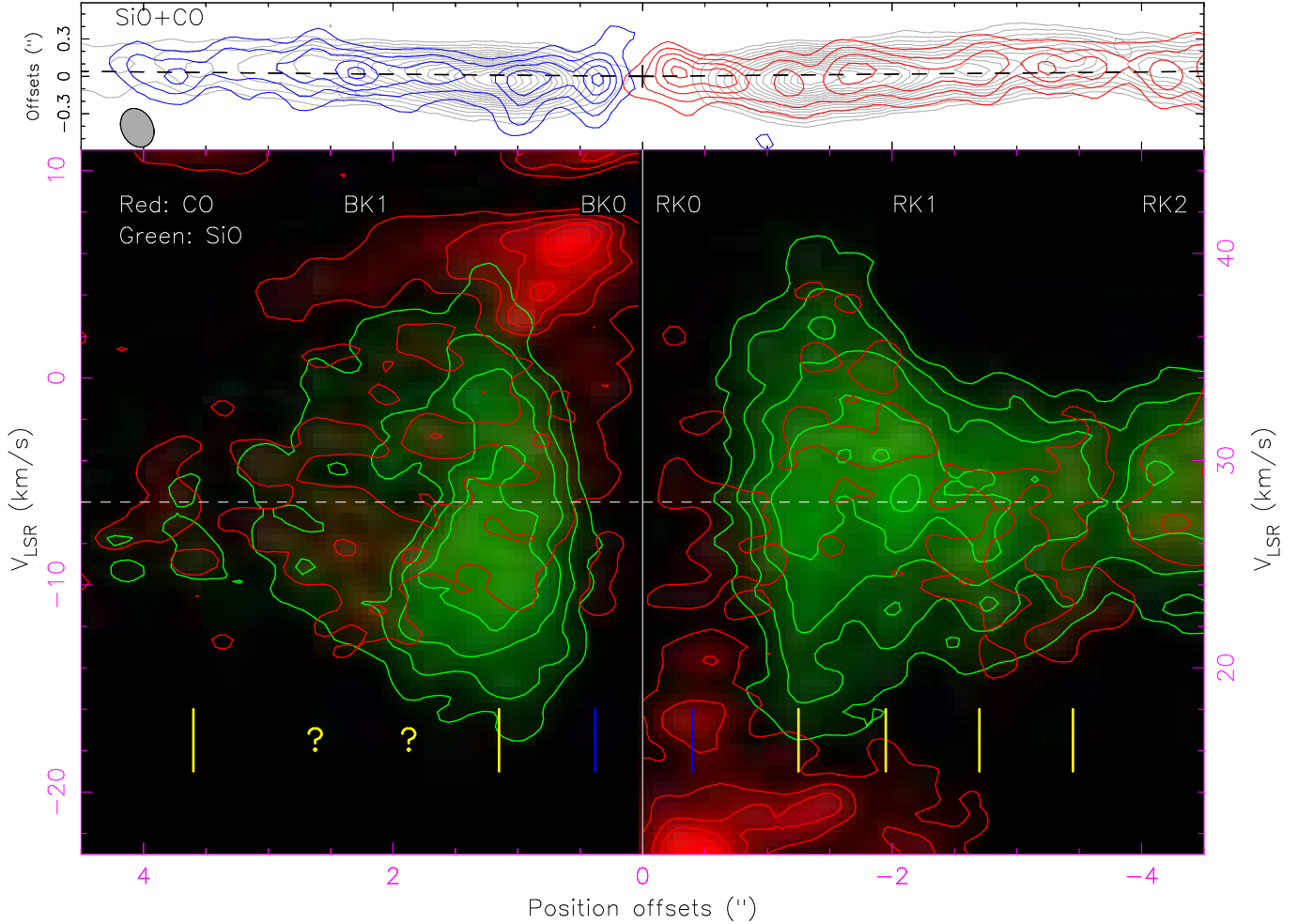

\centering
\putfig{0.7}{270}{f2.ps}
\figcaption[]
{
PV diagram cut along the jet axis in CO (red image and contours) for the two
inner pairs of knots, e.g., BK0, BK1, RK0 and RK1, in comparison to that in
SiO (green image and contours). As in \citet{Lee2009HH211},
the yellow lines mark the positions of the
sub-knots in knots BK1 and RK1, and the question marks indicate the possible
positions of the two sub-knots in knot BK1.  The blue lines mark the
positions of knots BK0 and RK0. The systemic velocity is 9.2 \vkm{}. The
white horizontal dashed lines mark the mean velocities of the jet, which are
$-$6 \vkm{} on the blueshifted side and 28 \vkm{} on the redshifted side, as
given in \citet{Lee2009HH211}. The SiO contours start at 0.15 \Jyb{} with a
step of 0.15 \Jyb{}. The CO contours start at 0.1 \Jyb{} with a step of 0.1
\Jyb{}.
\label{fig:pvjeti}
}
\end{figure}

Figure \ref{fig:pvjeti} shows the position-velocity (PV) diagram 
cut along the jet axis in CO for the two
inner pairs of knots, e.g., BK0, BK1, RK0 and RK1, in comparison to that in
SiO.
The innermost pair of CO knots, BK0 and RK0, allow us to study the jet
kinematics closer to the source than the innermost pair of SiO knots, BK1
and RK1. They are seen associated with a broad range of velocities, tracing the internal shocks closer to the source than the
SiO knots. The four sub-knots of BK1 and RK1 seen in SiO with a range of
velocities \cite[see][]{Lee2009HH211} have possible counterparts in CO in
the PV diagram (Fig. \ref{fig:pvjeti}, marked with yellow lines and question
marks). The separation between the first SiO sub-knots and the innermost
pair of CO knots is $\sim$ \arcsa{0}{8} (or $\sim$ 220 AU), roughly the same
as that in between the sub-knots. This suggests that the innermost pair of
CO knots and the sub-knots of BK1 and RK1 are all the internal shocks
produced by the same mechanism, namely by a periodical variation in the jet
velocity as suggested in \citet{Lee2009HH211}. 



\begin{figure} [!hbp]
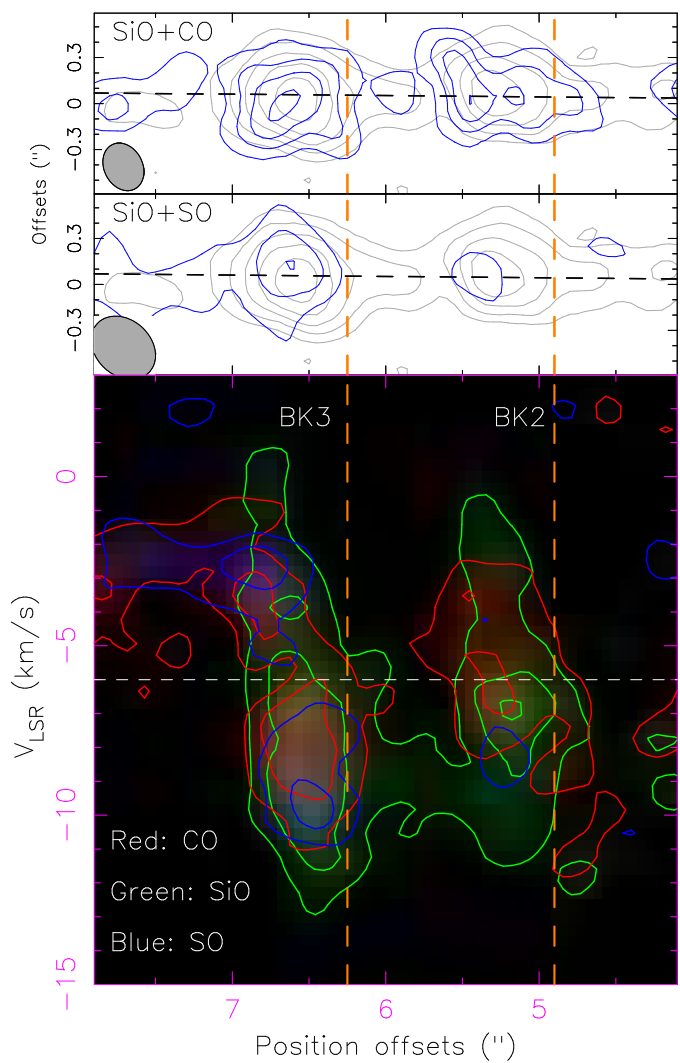

\centering
\putfig{0.75}{270}{f3.ps}
\figcaption[]
{
PV diagrams cut along the jet axis in CO (red image and contours), SiO
(green image and contours), and SO (blue image and contours) for knots BK2
and BK3. The white dashed line indicates the mean velocity for the eastern
component of the jet. The orange dashed lines separate the heads and tails
for knots BK2 and BK3, as in \citet{Lee2009HH211}. The SiO contours start at
0.15 \Jyb{} with a step of 0.15 \Jyb{}. The CO contours start at 0.1 \Jyb{}
with a step of 0.1 \Jyb{}. The SO contours start at 0.08 \Jyb{} with a step
of 0.08 \Jyb{}.
\label{fig:pvjetBK23}
}
\end{figure}

The knots BK2 and BK3 are better resolved, allowing us to study the detailed
kinematics in different emissions in the internal shocks. 
Figure \ref{fig:pvjetBK23} shows the PV diagrams cut along the jet axis in CO,
SiO, and SO for these knots.
In SiO, these
knots clearly show a head-tail morphology, with the heads associated with a
broad range of velocities and thus tracing the
internal shocks. With respect to the mean velocity of the jet reported in
\citet{Lee2009HH211}, the nearside of the heads is blueshifted and the
farside is redshifted, because of the sideways ejection of the jet material
in the shocks \cite[see also][]{Santiago2009}. In the heads of these knots,
the CO emission appears ahead of the SiO emission and the velocity range of
the CO emission is smaller than that of the SiO emission. Since SiO emission
traces denser and warmer regions than CO emission, these differences suggest
that the SiO emission traces stronger shocks closer to the shock front and
the CO emission traces the weaker shocks in the downstream.  As for the SO
emission, only knot BK3 is bright enough for kinematic study. In that knot,
the SO emission is slightly ahead of the SiO emission with a smaller
velocity range, tracing weaker shocks than the SiO emission. Also, the SO
emission seems slightly behind (i.e., upstream) the CO emission and 
peaks at higher velocity offsets from the mean jet velocity than
the CO emission, tracing stronger shocks than the CO emission.



\subsection{Column and Volume Densities of the jet}\label{sec:densities}



The column and volume densities of the jet can be estimated from the CO
emission of the two innermost pairs of knots (i.e., knots BK0, BK1, RK0, and
RK1). At the resolution of $\sim$ \arcsa{0}{46}$\times$\arcsa{0}{36}, the
mean intensity of the CO emission toward these knots is $\sim$ 3.2 \Jybk{}
(see Fig. \ref{fig:jet}b). The abundance of CO relative to H$_2$ is assumed
to be $\sim 4 \times 10^{-4}$ as if the CO gas is formed via gas-phase
reactions in an initially atomic jet \cite[][see also \S
\ref{sec:disabun}]{Glassgold1991}. The kinetic temperature
the CO emission can be assumed to be 200 K, slightly lower than the kinetic temperature 
derived from the
CO emission in much higher transitions in the far infrared, which was found
to be $\gtrsim$ 250 K \citep{Giannini2001}. 
Assuming LTE, the excitation temperature of the CO
emission can be assumed to be the same as the kinetic temperature.
Assuming that the CO emission
is optically thin, the column density of the jet in H$_2$ is $N\sim$ 5.5
$\times10^{20}$ \cms{}. However, the true column density of the jet is
actually higher because the jet is spatially unresolved perpendicular to the
jet axis. With the beam having a size $b \sim$ \arcsa{0}{45} or 125 AU
perpendicular to the jet axis and the jet having a deconvolved width
$\lesssim$ 40 AU \cite[derived from the SiO jet in][]{Lee2009HH211}, the
true column density of the jet is $N_\mathrm{a} \gtrsim N (125/40) \sim 1.7
\times 10^{21}$ \cms{}. As a result, the volume density of the jet is $n
\gtrsim N_\mathrm{a}/40\mathrm{AU} \sim 2.8 \times 10^6$ \cmc{}, about two
orders of magnitude higher than the critical density of CO (\COt{})
line, which is $3\times10^{4}$ \cmc{} at 200 K. On
the other hand, the volume density derived here is one order of magnitude
lower than the critical density of the SO line, and two orders of magnitude
lower than that of the SiO line at 200-500 K. Note that the critical densities here
are derived using the collisional rate coefficients listed
in the Leiden Atomic and Molecular database \citep{Schoier2005}.

\subsection{SO and SiO abundances in the jet} \label{sec:abundance}

The abundances of SO and SiO relative to H$_2$ can be estimated by dividing
SO and SiO column densities by that of H$_2$ derived from the CO emission,
respectively. However, since the SO and SiO emissions trace stronger
shocks and thus denser material than the CO emission, the abundances
estimated here are likely to be the upper limits.

At the resolution of \arcsa{0}{46}$\times$\arcsa{0}{36}, the mean SO
intensity toward the innermost pair of SO knots (i.e., knots BK1 and RK1) is
$\sim$ 1 \Jybk{} (see Fig. \ref{fig:jet}a). The excitation temperature of 
the SO emission is assumed to be the same as that of the CO emission, 
but it
could be higher because the SO emission seems to trace stronger shocks than
the CO emission as seen in knot BK3 in the downstream (see \S
\ref{sec:kinematics}).
Assuming that the SO emission
is optically thin, the SO column density is $\sim 1.1\times10^{15}$ \cms{}.
Therefore, the abundance of SO is $\sim$ 2$\times 10^{-6}$. It is $\sim$ 400
times that in cold quiescent clouds (cores), e.g., TMC-1, which is $\sim 5
\times 10^{-9}$ \citep{Ohishi1998}.




At the resolution of \arcsa{0}{46}$\times$\arcsa{0}{36}, the SiO intensity
toward the innermost pair of SiO knots decreases rapidly with the distance
from the source from $\sim$ 20 to 5 \Jybk{} (Fig. \ref{fig:jet}a). Assuming
an excitation temperature of 500 K and optically thin emission as in
\citet{Lee2009HH211}, the SiO column density at the SiO knots is
(4.4--1.0)$\times 10^{15}$ \cms{}, resulting in a SiO abundance of
(8--2)$\times10^{-6}$. Therefore, the SiO abundance is highly enhanced by
about five orders of magnitude over that in quiescent molecular clouds where
the abundance of SiO is $\lesssim 10^{-11}$ \citep{Ziurys1989}.

\section{An orbiting source jet model for the wiggle}
\label{sec:model}

\begin{figure} [!hbp]
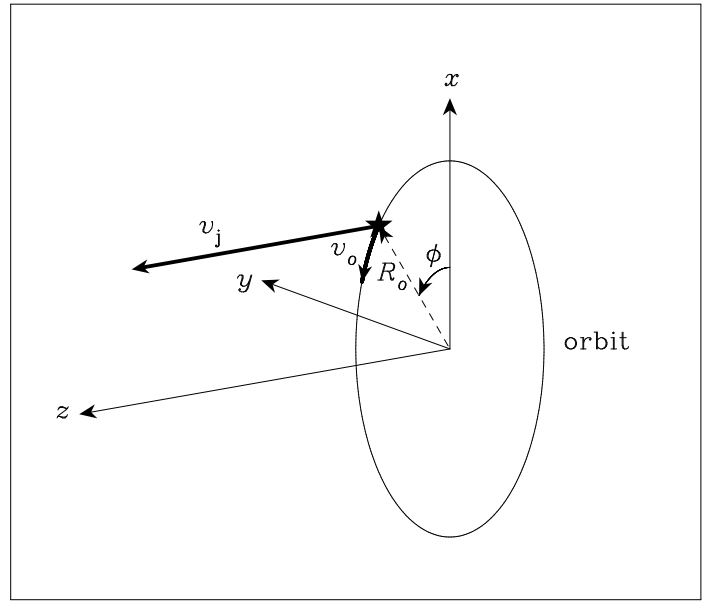

\centering
\putfig{0.5}{270}{f4.ps}
\figcaption[]
{
Schematic diagram of the orbiting source jet model. Star marks the source
position in the orbit.
\label{fig:jetmodel}
}
\end{figure}
\begin{figure} [!hbp]
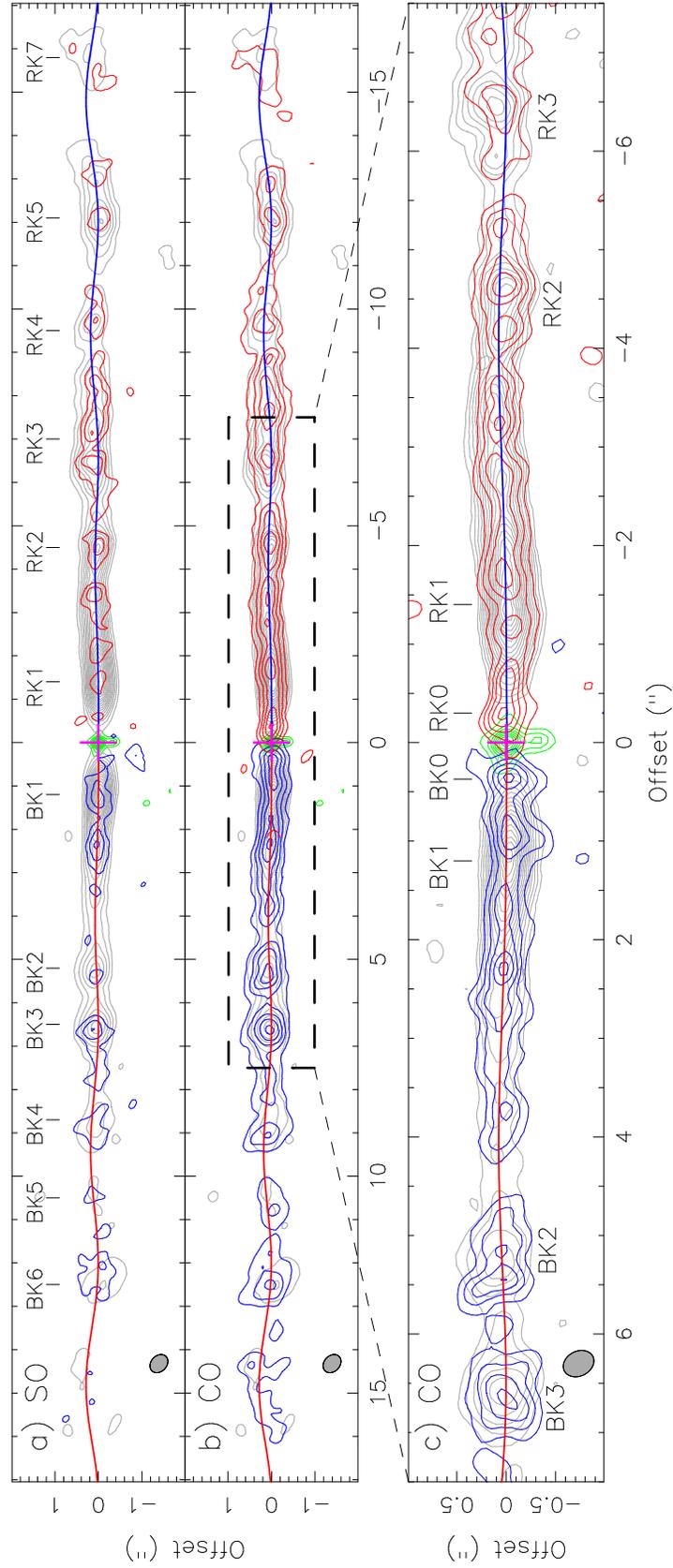

\centering
\putfig{0.925}{0}{f5.ps}
\figcaption[]
{
The orbiting source jet model for the wiggle plotted on top of Figure
\ref{fig:jet}. The red and blue curves are derived from our best-fit model
(see the text in \S \ref{sec:model}).
\label{fig:jetfit}
}
\end{figure}

The reflection-symmetric wiggle of the jet could be due to an orbital motion
of the jet source in a binary system \citep{Fendt1998,Masciadri2002}.  If
that is the case, then by modeling the wiggle of the jet, we could obtain
the parameters of the orbital motion and then discuss the nature of the
source (see \S \ref{dis:binary}). To model, we adopt the right-handed
Cartesian coordinate system as shown in Figure \ref{fig:jetmodel}, with the $+z$-axis aligned with the eastern
component of the jet. For simplicity, the jet
source is assumed to have a circular orbit in the $x$-$y$ plane, with the
jet propagating ballistically in the $z$-axis at a velocity of $\eqt{v}{j}$. 
The orbit has a period $\eqt{P}{o}$, an orbital velocity $\eqt{v}{o}$, and a
radius $\eqt{R}{o}$. Assuming that the orbital motion of the jet source is
in the same direction as the rotation motion of the \HCOP{} envelope-disk
\citep{Lee2009HH211}, we have $\eqt{v}{o}>0$ (i.e., positive for
right-handed orbital motion). Then for a jet lying close to the plane of the
sky as in the case of HH 211, the trajectory of the jet in the plane of the
sky can be approximately given by
\begin{eqnarray}
x &=& -\sqrt{\eqt{R}{o}^2 + (\eqt{v}{o}\frac{z}{\eqt{v}{j}})^2} 
\cos[\eqt{\phi}{0}-\frac{\eqt{v}{o}}{\eqt{R}{o}}\frac{z}{\eqt{v}{j}}+\tan^{-1}
    (\frac{\eqt{v}{o}}{\eqt{R}{o}}\frac{z}{\eqt{v}{j}})]\nonumber \\
&&+ \eta\; z
\end{eqnarray}
where $\phi_0$ is the current phase angle of the jet source in the orbit
measured from the $+x$-axis toward the $+y$-axis direction, and $\eta$ is to
account for the small-scale bending of the jet to the northeast
\citep{Lee2009HH211}. Define the periodic length as
\begin{equation}
\triangle z \equiv \eqt{v}{j} \eqt{P}{o}
\end{equation}
and the velocity ratio as
\begin{equation}
\kappa \equiv \frac{\eqt{v}{o}}{\eqt{v}{j}}, 
\end{equation}
we have
\begin{equation}
\eqt{R}{o} = \frac{\eqt{v}{o} \eqt{P}{o}}{2 \pi} 
= \frac{\kappa \triangle z}{2 \pi}
\end{equation}
and thus
\begin{eqnarray}
x &=& -\sqrt{(\frac{\kappa \triangle z}{2 \pi})^2 + (\kappa z)^2} 
\cos(\eqt{\phi}{0}-\frac{2 \pi z}{\triangle z}+\tan^{-1}
    \frac{2 \pi z }{\triangle z})\nonumber \\
&&+ \eta\; z
\end{eqnarray}
When $z \gg \frac{\triangle z}{2\pi}$, we have
\begin{eqnarray}
x \approx -\kappa z
\cos(\eqt{\phi}{0}-\frac{2 \pi z}{\triangle z}+\frac{\pi}{2})+\eta \;z
\label{eq:xbendm}
\end{eqnarray}
which means that the wiggle due to the orbital motion of the jet source 
has an opening angle $\alpha \approx 2\kappa$.
Figure \ref{fig:jetfit} shows the best fit of this model to the SiO, CO, and SO data simultaneously.
As can be seen, the wiggle of the jet 
can be reasonably fitted by this model.
The best-fit parameters are $\triangle z=1530\pm25$ AU ($\sim$
\arcsa{5}{46}$\pm$\arcsa{0}{09}),
$\kappa=0.0094\pm0.0007$ (or 0.54\degree{}$\pm$0.04\degree{}), 
$\eqt{\phi}{0} =155$\degree{}$\pm$13\degree{},
and $\eta=0.0098\pm0.0009$ (or 0.56\degree{}$\pm$0.05\degree{}). 
Thus, $\eqt{R}{o}\sim2.3\pm0.2$ AU, $\eqt{v}{o}\sim1.60\pm0.56$ \vkm{}, 
and $\eqt{P}{o} \sim 43\pm23$ yrs,
assuming that $\eqt{v}{j}=170\pm60$ \vkm{} as in \citet{Lee2009HH211}.
In this model, the wiggle of the jet has an opening angle of
$\alpha \sim$ 1.1\degree{}, similar to that seen in the observations (Fig.
\ref{fig:jetfit}). About 3
cycles of wiggle are seen on each side of the source.
The specific angular momentum for the orbital motion is $\sim$ 3.7$\pm1.3$ AU
\vkm{}, comparable to that of the jet rotation yet to be confirmed, which is
$\lesssim$ 5 AU \vkm{} \citep{Lee2009HH211}.

\section{Discussion}

\subsection{SMM1: A very low-mass protobinary?}\label{dis:binary}


It has been argued that almost all stars must form in binary or multiple
systems because of the fragmentation of star-forming dense molecular cores
\citep{Goodwin2007}. Could the source SMM1 itself be a protobinary in the
Class 0 phase (i.e., the early phase of star formation) since the wiggle is
reasonably reproduced by an orbiting source jet model? In order to
investigate this possibility, we evaluate here the possible mass of the
source, assuming that the source is a binary. Let $M_j$ be the mass of the
jet source, $M_c$ be the mass of the companion, and $m$ be the mass ratio of
the jet source to the companion, i.e., $M_j=m M_c$, then the binary
separation is $a=(1+m)\eqt{R}{o}$. From the Kepler's 3rd law of orbital
motion, the total mass of the binary in terms of the model parameters is
then 
\begin{eqnarray} 
M_t &\approx & 9.5\times 10^{-4}\ (1+m)^3 \nonumber \\
&&\times (\frac{\eqt{v}{j}}{100 \;\vkme{}})^2 (\frac{\kappa}{1^\circ})^3
\frac{\triangle z}{100 \;\rm{AU}} \;\;M_\odot 
\end{eqnarray} 
\mbox{}\\
With the best-fit values of $\kappa$ and $\triangle z$, and the jet velocity
$\eqt{v}{j}=170$ \vkm{}, we have 
\begin{eqnarray} 
M_t &\approx & 6.9 \; (1+m)^3 \;\;M_\textrm{\scriptsize Jup} 
\end{eqnarray} 
To determine the value of $m$, we resort to the previous estimation of the
mass of the source. Previously, the mass of the source has been estimated to
be $\sim$ 60 \Jupmass{} from an evolutionary model \citep{Froebrich2003}. 
This mass is also consistent with that derived from the \HCOP{} rotating
envelope-disk, assuming that the rotation is Keplerian \citep{Lee2009HH211}.
If we adopt this mass as the total mass of the binary, then $m\approx1$.
This value of $m$ is in good agreement with that found in the binaries of
very low-mass stars and brown dwarfs \citep{Close2003,Bouy2006}, which tend to have
equal mass companions. In addition, the binary separation would be
$2\eqt{R}{o}\sim$ 4.6 AU, also consistent with that found in those binaries
\citep{Close2003,Bouy2006}, which have a range of separation between
0--30 AU. Therefore,
SMM1 itself could indeed be a very low-mass protobinary because of the
fragmentation at the beginning of star formation \cite[see,
e.g.,][]{Machida2008bin}. It is unclear, however, why the jet source, with
only $\sim$ 30 \Jupmass{}, can have such an energetic collimated jet, but
the companion as massive as the jet source does not have a jet or an
outflow. It is also unclear if this protobinary will remain substellar in
the final stage since it is surrounded by a compact envelope with an
estimated mass of $\sim$ 50 \Jupmass{} \cite[although it is a lower
limit, see][]{Lee2007HH211} at the eastern edge of the ammonia envelope
\citep{Lee2009HH211}? Also, since the source SMM1 could form a triple system
with the source SMM2 detected at $\sim$ 84 AU to the southwest with a
planetary mass \citep{Lee2009HH211}, could the source SMM2 be ejected from
the SMM1 binary system in the early stage of star formation because of
dynamical decay \citep{Goodwin2007}?
The HH211, if indeed a       
triple system consisting of a close binary ($<$30 AU) 
with a third component at $\sim$ 100 AU, would be quite similar to those found
in Upper Scorpius by \citet{Bouy2006}.



\subsection{The mass-loss rate and accretion rate}

Here we estimate the mass-loss rate of the jet and the accretion rate
toward the central source, in order to compare their ratio with that predicted in
the current jet launching models.
The (two-sided) mass-loss rate of the jet can be given by
\begin{equation}
\dot{M}_j \sim 2 v_j m_{\textrm{\scriptsize H}_2} N b
\end{equation}
where $N$ and $b$ are the column density of the jet and the
beam size perpendicular to the jet axis, respectively, as given in \S\ref{sec:densities},
and $m_{\textrm{\scriptsize H}_2}$ is the mass of an H$_2$ molecule.
With the jet velocity $v_j \sim$ 170 \vkm{}, the mass-loss 
rate is $\dot{M}_j \sim 1.8 \times10^{-6}$ \smassrate{}, as found in
\citet{Lee2007HH211}.
The mechanical luminosity of the jet is then
\begin{equation}
L_j = \frac{1}{2}\dot{M}_j v_j^2 \sim 4.3 L_\odot
\end{equation}
roughly the same as the bolometric luminosity of the source, which is
$L_\textrm{\scriptsize bol} \sim 3.6 L_\odot$ \citep{Froebrich2005}. 
The jet is likely powered by accretion as in the current jet launching
models \cite[see, e.g.,][]{Shu2000,Pudritz2007}, and our result suggests that
the jet can indeed carry away a large fraction of the accretion power 
in the early phase
of star formation as pointed out by \citet{Cabrit2000}. Therefore, the
accretion rate should be
estimated using both the bolometric luminosity of the source and the
mechanical luminosity of the jet. Assuming
that the source SMM1 is a single protostar with a stellar mass of $M_\ast
\sim 60$ \Jupmass{} \citep{Froebrich2003,Lee2009HH211} and a stellar radius
of $R_\ast \sim 2 R_\odot$ \citep{Stahler1988,Machida2008rad}, then the
accretion rate is $\dot{M}_a \sim (L_\textrm{\scriptsize bol}+L_j) R_\ast/GM_\ast \sim 8.5
\times 10^{-6}$ \smassrate{}. Thus, the mass-loss rate is estimated to be
$\sim$ 20\% of the accretion rate, consistent with that predicted in the
current MHD models \citep{Shu2000,Pudritz2007}.  However, the source SMM1
itself could be a binary (see \S \ref{dis:binary}). If the jet source has a
mass of $\sim$ 30 \Jupmass{} and it is responsible for half of the
bolometric luminosity, then the accretion rate toward the jet source is
$\sim 1.3\times 10^{-5}$ \smassrate{}. In this case, the mass-loss rate is
$\sim$ 14\% of the accretion rate, also consistent with that derived simply
from the law of energy conservation and the jet's angular momentum reported
in \citet{Lee2009HH211} \cite[e.g.,][]{Soker2009}. Further observations are
really needed to constrain our estimation.

Now let us check if the accretion rate is acceptable by comparing the
accretion time with the outflow dynamical time. If the accretion rate was
the same in the past, the accretion time would be $M_\ast/\dot{M}_a$, i.e.,
$\sim$ 7000 yrs if the source is single, or $\sim$ 2300 yrs
if the source is a binary. The outflow dynamical time can be estimated with
the tip of the eastern outflow lobe in H$_2$ \citep{Hirano2006} because the
bright H$_2$ emission there requires the shock velocity there to be
$\lesssim$ 55 \vkm{}, otherwise H$_2$ there will be all dissociated
\citep{Smith1991,OConnell2005}. Therefore, with the tip at $\sim$ 12500 AU from the
source, the dynamical time for the outflow is $\gtrsim$ 1100 yrs. As a
result, the accretion time is not shorter than the outflow dynamical time,
and thus the accretion rate is acceptable.  
The accretion rate is high
probably because of the extremely young age of the source SMM1.
Note that, if the source SMM1 is a binary, then in our estimation the
accretion time is only two times the outflow dynamical time. If that is the
case, the jet could be launched soon after the accretion started.



\subsection{Origin of SO and SiO emission: shock
enhancement}\label{sec:disabun}





The abundances of SiO and SO in the gas phase are found to be highly
enhanced in the jet as compared to the quiescent molecular clouds, even
close to within 300 AU from the source where the dynamical time scale is
$<10$ yrs. Here we discuss the possible origins of the abundance enhancement
of these molecules.

Although SiO abundance enhancement has been found in many Class 0 jets, its
origin is still not well determined.  In HH 211, the abundance enhancement
of SiO is closely related to the shocks, with the emission associated with a
broad range of velocities. 
As pointed out by Schilke et al. (1997) in their early work, the
     shocks are likely required to generate the SiO gas by
     either (1) a release of SiO from dust grains, or (2) a release of
     Si atoms followed by conversion to SiO through gas-phase chemistry.
The jet indeed seems to be dusty, with the 352 GHz continuum emission extending
along the jet axis from the source \citep{Lee2009HH211}.
Recently, \citet{Gusdorf2008a} have studied the formation time of the SiO from Si
via gas-phase reactions in stationary C-type shocks with 
grain sputtering of Si-bearing material for the preshock number
density of hydrogen nuclei $\nH=10^4 - 10^6$ \cmc{}.
From their results,
if the shock velocity is greater than 30 \vkm{},
we can approximate the SiO formation time to be roughly given by $10^7/\nH$ yrs.
If we extrapolate that SiO formation time for 
HH 211 that has $\nH = 2 n \gtrsim 6\times
10^6$ \cmc{} (see \S \ref{sec:densities} for $n$), then the formation
time is $\lesssim$ 2 yrs for HH 211, short enough to produce the SiO emission in the SiO
knots.
Note that non-stationary C-type shocks \citep{Gusdorf2008b} and 
J-type shocks with more sophisticated grain processes \citep{Guillet2009}
could also be responsible for the SiO emission and more studies are
needed to explore those possibilities.
There is a lack of SiO
emission [not only in J=8-7 transition, but also in lower transitions in
J=5-4 (Gueth et al. in prep) and J=1-0 \citep{Chandler2001}] at the base of
the jet in the innermost pair of CO knots, BK0 and RK0.  This lack of SiO
emission is unlikely due to a lack of material because the jet is expected
to have the same density along the jet axis except near the launching point
where the density is much higher \citep{Shu2000}. This lack of SiO emission
suggests that the SiO is not abundant there in the gas phase. If the SiO
were there abundant in the gas phase, the shocks that excite the CO emission
\cite[which was also detected in higher transition in the far infrared
by][]{Giannini2001} would also excite the SiO emission. 



SO abundance enhancement has also been seen toward other Class 0 sources,
e.g., L1448, HH 212, and L1157, because of shocks. 
In L1157, the SO emission
is seen in the prominent bow shocks at $\gtrsim$ 20000 AU away from the
source, but no jetlike SO emission is seen around the source
\citep{Bachiller2001}. 
In L1448, the SO emission is
seen at shock-precursor and post-shock velocities \citep{Jimenez2005}.
In HH 212, the SO emission at the base is
jetlike as in HH 211 and it extends to $\sim$ 800 AU from the source
\citep{Lee2007HH212}. Note that in that source SO emission in lower transition has also
been detected upto 8000 AU to the south along the jet \citep{Lee2006}.
Thus, the abundance of SO in the jet seems to decrease
from the younger sources, e.g., HH 211 and HH 212, to the older sources,
e.g, L1157 (which shows no clear jetlike SO emission), like that of SiO
\citep{Shang2007}.
In L1157, it was proposed that sulfur is
released from dust grains in the form of H$_2$S in the shocks and that is
then oxidized to SO \citep{Bachiller2001}.    In L1448, H$_2$S
emission is present in the shock-precursor, but is missing in the post-shock
gas. 
\citet{Jimenez2005} considered oxidation of H$_2$S to SO in shock, but
judged the timescale [100 -- 1000 yrs suggested by \citet {Wakelam2004}] too
long.  They thus proposed that the SO molecules might be abundant on dust grains and directly released from
grains in the shocks.
HH 211 and HH 212 could have the same
origin of abundance enhancement in SO. In HH 211, the shocks can release
sulfur from grains in the form of e.g., H$_2$S, and then oxidize it to SO,
as proposed in L1157.
If the jet is rich in atomic oxygen (OI) and carbon ion (CII) \cite[as can
be seen in ISO observations in][] {Giannini2001}, then oxidation of H$_2$S
to SO could take only a few years.  
For example, the following sequence would have timescales
less than 5 yrs in a shock environment with $\nH \sim 10^6$ \cmc{},
$T_\textrm{\scriptsize gas} \sim 1000$ K, $n$(HI)/$n$(H$_2$) $\sim$ 0.01, 
and $n$(OI)/$\nH$ $\sim 10^{-4}$:
H$_2$S + H $\rightarrow$ HS + H$_2$; HS + O $\rightarrow$ SO + H. 
In this example,
the rate coefficients are 2.5$\times 10^{-11}$ \cmc{} s$^{-1}$ and
1.3$\times 10^{-10}$ \cmc{} s$^{-1}$ \cite[RATE06 by][]{Woodall2007}, and
the conversion timescales for H$_2$S $\rightarrow$ HS and HS $\rightarrow$ SO are
0.13 yrs and 3 yrs, respectively.  (The timescale for
HS + H $\rightarrow$ S + H$_2$ is 0.13 yrs, but the timescale for reverse
S + H$_2$ $\rightarrow$ HS + H is 2.6 yrs at 1000 K.  Thus, about 5 percent
of sulfur atoms will be in HS in 3 yrs and be available for
HS + O.)
Again, there is a lack of SO emission at the base of the jet in the
innermost pair of CO knots. This suggests that the SO is not abundant there,
neither in the gas phase nor on the dust grains. If the SO were there on the
dust grains, it would evaporate into the gas phase and then be excited to
produce the SO emission.  

We now speculate what could cause the lack of SiO and SO abundances in the
gas phase at the base of the jet in the innermost pair of CO knots. 
Since the
mass-loss rate in the jet is high ($\sim$ $1.8\times10^{-6}$ \smassrate{}),
SiO and CO as well could have formed via gas-phase reactions in an initially
atomic jet close to the launching point ($<$ 0.1 AU) \citep{Glassgold1991}.
If that is the case, the lack of gas-phase SiO in the innermost pair of CO
knots suggests that the SiO could be afterward dissociated e.g., by
the possible presence of a far-UV radiation field near the source. 
However, it is also possible that the SiO is rapidly depleted onto
the grains or converted to SiO$_2$ \citep{Gusdorf2008a}
if the temperature of the jet indeed drops rapidly with the distance from
the launching point as predicted by \citet{Glassgold1991}.
In addition, the lack of
gas-phase SO in the innermost pair of CO knots also suggests that the SO, if
formed at the base, is afterward destroyed. On the other hand, the CO could
stay in the gas phase, giving rise to the CO emission even in the innermost
pair of CO knots. Both the SiO and SO in the gas phase can form again in the
shocks, as discussed above, producing the SiO and SO in the downstream. The
SO and SiO abundances in the jet are seen decreasing with the outflow
dynamical age, probably because the mass-loss rate decreases with time and so
does the formation rate of the SiO and SO molecules. 



\section{Conclusions}

SiO, CO, and SO can all be a good tracer of the HH 211 jet, tracing the
internal shocks in the jet. Although the SiO, CO, and SO emissions
show roughly the same morphology of the jet, there are detailed differences.
In particular, the CO emission traces the jet closer to the source than the
SiO and SO emissions. 
In addition, in the better resolved internal shocks, the CO emission is seen
slightly ahead of the SiO emission, likely because the SiO emission traces
stronger shocks closer to the shock front and the CO emission traces the
weaker shocks in the downstream.  In the internal shock where the SO
emission is bright, the SO emission appears slightly ahead of the SiO
emission, tracing weaker shocks than the SiO emission.

The jet is seen with more than one cycle of wiggle on both sides of the
source SMM1 in SiO, CO, and SO. The wiggle is reflection-symmetric about the
source and can be reasonably fitted by an orbiting source jet model. The
best-fit parameters suggest that the source itself could be a very low-mass
protobinary with a total mass of $\sim$ 60 \Jupmass{}. The protobinary may
have two equal-mass sources and a binary separation of $\sim$ 4.6 AU, in
good agreement with those found in the binaries of very low-mass stars and
brown dwarfs.

The abundances of SiO and SO in the gas phase are found to be highly
enhanced in the jet as compared to the quiescent molecular clouds, even
close to within 300 AU from the source where the dynamical time scale is
$<10$ yrs. The abundance enhancements of these molecules are closely related
to the internal shocks in the jet. 
The detected SiO is either the consequence of
the release of Si-bearing material
from dust grains or of its formation via gas chemistry in the shocks.
The SO, on the other
hand, seems to form via gas chemistry in the shocks.


\acknowledgements
We thank the SMA staff for their efforts in running and maintaining the
array, and the anonymous referee for the precious comments.
C.-F. Lee thanks Frank Shu for fruitful conversations and Ngoc Phan-Bao for
useful discussion. A.P. is
grateful to Joaqu\'{i}n Santiago-Garc\'{i}a, Robert Estalella and Nuria
Hu\'elamo for useful discussions.


\centering
\putfig{0.925}{0}{f1.ps}
\figcaption[]
{
SO and CO maps of the jet on top of the SiO map (gray contours) of the jet
and the 352 GHz continuum map (green contours) of the envelope-disk from
\citet{Lee2009HH211}. The maps have been rotated by 63.4\degree{}
counterclockwise. 
The blueshifted and redshifted SiO emission are integrated from -21.2 to 9.2 \vkm{}
and from 9.2 to 47.5 \vkm{}, respectively.
The cross marks the position of the source SMM1. The knots
have the same notations as in \citet{Lee2009HH211}. \tlabel{a} shows the
blueshifted and redshifted SO emission, integrated from -16 to 9.2 \vkm{}
and from 14.5 to 43.5 \vkm{}, respectively.  No SO emission is detected 
from 9.2 to 14.5 km/s. The contours start at 0.52
\Jybk{} with a step of 0.52 \Jybk{}. The SiO contours start at 1 \Jybk{}
with a step of 1.5 \Jybk{}. \tlabel{b} and \tlabel{c} show the blueshifted
and redshifted CO emission, integrated from -16 to 0 \vkm{} and from 20 to
43 \vkm{}, respectively. In \tlabel{b}, the CO contours start at 0.56
\Jybk{} with a step of 0.84 \Jybk{}. The SiO contours are the same as in
\tlabel{a}. In \tlabel{c}, the CO contours start at 0.5 \Jybk{} with a step
of 0.5 \Jybk{}. The SiO contours start at 0.84 \Jybk{} with a step of 0.84
\Jybk{}. The synthesized beams for the lines are
\arcsa{0}{46}$\times$\arcsa{0}{36} (natural weighting) in \tlabel{a} and
\tlabel{b}, and \arcsa{0}{35}$\times$\arcsa{0}{25} (super-uniform weighting)
in \tlabel{c}.
\label{fig:jet}
}


\end{document}